\documentclass[aps,twocolumn,prl,superscriptaddress,showpacs,10pt]{revtex4-1}
\usepackage{amsmath} \usepackage{amsfonts} \usepackage{amssymb}
\usepackage{amsthm} \usepackage{mathtools} \usepackage{mathrsfs}
\usepackage{txfonts} \usepackage{color} \usepackage{xspace}
\usepackage{epsfig} \usepackage{graphicx} \usepackage[colorlinks,
  urlcolor=black, citecolor=blue, link color=black]{hyperref}

\def \Re{\text{Re}}

\def\Dbar {\kern 0.2em\bar{\kern -0.2em D}{}\xspace}

\def\Dzb   {\ensuremath{\Dbar^0}\xspace}
\def\DzDzb {\ensuremath{D^0 {\kern -0.16em \Dzb}}\xspace}
\def\Dp    {\ensuremath{D^+}\xspace}

\def\Dzzb {\kern 0.2em\overset{(\bar{\kern -0.2em D})}{}\xspace}

\def\Dzbs   {\ensuremath{\Dbar^{*0}}\xspace}
\def\DzsDzbs {\ensuremath{D^{*0} {\kern -0.16em \Dzbs}}\xspace}

\def\Bbar  {\kern 0.18em\bar{\kern -0.18em B}{}\xspace}

\def\Bzb   {\ensuremath{\Bbar^0}\xspace}
\def\BzBzb {\ensuremath{B^0 {\kern -0.16em -\Bzb}}\xspace}

\def\Kbar  {\kern 0.2em\bar{\kern -0.2em K}{}\xspace}

\def\Kp    {\ensuremath{K^+}\xspace}
\def\Km    {\ensuremath{K^-}\xspace}

\def\Bbar  {\kern 0.18em\bar{\kern -0.18em B}{}\xspace}

\def\Bzb   {\ensuremath{\Bbar^0}\xspace}
\def\BzBzb {\ensuremath{B^0 {\kern -0.16em -\Bzb}}\xspace}

\def\nn    {\nonumber}
\def\sss{\scriptscriptstyle}

\def\pip    {\ensuremath{\pi^+}\xspace}
\def\pim    {\ensuremath{\pi^-}\xspace}

\def\piz    {\ensuremath{\pi^0}\xspace}

\newcommand{\modulus}[1]{\left| #1 \right|}

\def\sss{\scriptscriptstyle}

\def\bBar#1{\hbox{$#1$\kern -1.85em\raise1.6ex\hbox{{\raise.35ex
\hbox{$~{\sss(}$}}$-${\raise.35ex\hbox{${\sss )}$}}}\kern 0.3em}}

\allowdisplaybreaks

\begin{document}

\title{On testing violations of Bose and
\texorpdfstring{$CPT$}{CPT} symmetries via Dalitz plots and Dalitz
`prism'}

\author{Dibyakrupa Sahoo}
\affiliation{The Institute of Mathematical Sciences, Taramani,
Chennai 600113, India}

\author{Rahul Sinha}
\affiliation{The Institute of Mathematical Sciences, Taramani,
Chennai 600113, India}

\author{N.~G.~Deshpande}
\affiliation{Institute of Theoretical Science, University of Oregon,
Eugene, OR 94703, USA}

\date{\today}

\begin{abstract}
Bose symmetry and $CPT$ symmetry are two very fundamental symmetries
of Nature. However, the validity of these symmetries in diverse
phenomena must be verified by experiments. We propose new techniques
to probe these two fundamental symmetries in the realm of mesons by
using the Dalitz plot of a few three-body meson decays. Since these
symmetries are very fundamental in nature, their violations, if any,
are expected to be extremely small. Hence, observing their violations
requires study of a huge data sample. In this context we introduce a
new three-dimensional plot which we refer to as the Dalitz
`prism'. This provides an innovative means for acquiring the huge
statistics required for such studies. Using the Dalitz plots and the
Dalitz prisms we chart out the way to probe the violations of Bose and
$CPT$ symmetries in a significant manner. Since mesons are unstable
and composite particles, testing the validity of Bose symmetry and the
$CPT$ symmetry in these cases are of paramount importance for
fundamental physics.
\end{abstract}


\maketitle

The statement that a state made up of two identical bosons does not
alter under exchange of the two bosons is the dictum of Bose
symmetry~\cite{ref:Bose-1924}. This along with the Fermi
statistics~\cite{ref:Fermi-1926} forms one of the cornerstones of
modern physics, the famous spin-statistics theorem. Within the
conventional Lorentz invariant and local quantum field theory, even a
small violation of Bose symmetry is impossible. There have been
therefore a lot of interest in experiments looking for Bose symmetry
violation as a means of testing the present theoretical
framework. Theoretical ideas and experimental investigations for Bose
symmetry violations have looked at the spin-0 nucleus of oxygen
${}^{16}\text{O}$~\cite{ref:Hilborn, ref:Tino-1994}, molecules such as
${}^{16}\text{O}_2$ and $\text{CO}_2$ \cite{ref:Hilborn-1996,
  ref:Angelis, ref:Modugno-1998, ref:Tino-2000}, photons
\cite{ref:Fivel-1991, ref:Gerry-1997, ref:Ignatiev, ref:manko,
  ref:DeMille-1999, ref:DeMille-2000}, pions
\cite{ref:Greenberg-1988um} and Bose symmetry violating transitions
\cite{ref:Mills, ref:Gidley, ref:Asai, ref:OPAL, ref:English-2010,
  ref:Kozlov-2012ag, ref:Gninenko-2011ws}. Theoretically a scenario
where Bose symmetry is not exact swings open doors to a plethora of
avenues for new physics \cite{ref:Jackson-2008bs, ref:Jackson-2008xq,
  ref:Greenberg-1991av, ref:Greenberg-1991au,
  ref:Greenberg-2000zy}. Like the Bose symmetry, the very nature of
Lorentz invariant local quantum field theory encompasses another
fundamental symmetry of Nature, namely the $CPT$ symmetry. This
symmetry combines the operations of charge conjugation ($C$), parity
($P$) and time reversal ($T$). In the conventional settings of quantum
field theory, the $CPT$ symmetry is very closely related to both
spin-statistics theorem and Lorentz invariance~\cite{ref:Belinfante,
  ref:Pauli-Belinfante, ref:Pauli-1940, ref:Schwinger-1951,
  ref:Schwinger-1953, ref:Luders-1954, ref:Bell-1955, ref:Pauli-1955,
  ref:Luders-1957, ref:Jost-1957, ref:Jost-1960,
  ref:Streater-Wightman, ref:Fainberg-1969, ref:Fonda-Ghirardi,
  ref:Haag-1996, ref:Luders-1957pr, ref:Luders-1958,
  ref:streater-1980, Dalitz:1990md, ref:Greenberg-2002}. However,
$CPT$ invariance and the spin-statistics theorem need not be
connected~\cite{Dalitz:1990md, ref:Oksak-1968, ref:Bogoliubov-1975},
and there are examples of quantum field theories in the
literature~\cite{Majorana:1932rj, ref:Nambu-1966, ref:Nambu-1967} that
explicitly violate the $CPT$ invariance. Under $CPT$ transformation, a
particle becomes its antiparticle and vice versa with the same
three-momentum but with its helicity reversed. The $CPT$ invariance
also implies that a particle and its antiparticle must have the same
mass, decay width and lifetime. It is important to note that if $CPT$
invariance holds good but $CP$ is violated, then partial rate
asymmetries for a particle and its antiparticle can be different while
keeping their total decay rates
unchanged~\cite{ref:Okubo-1958}. Similarly, the $CPT$ invariance also
implies that the total scattering cross-section of two particles would
be equal to that of their antiparticles, but the partial scattering
cross-sections need not be equivalent if $CP$ is
violated~\cite{ref:Okubo-1962}. Though $CPT$ invariance is in concord
with our present theoretical framework of Standard Model of particle
physics, it needs to be thoroughly tested experimentally. The
literature is replete with many tests for $CPT$ violation, such as in
anomalous magnetic moments~\cite{Bluhm:1997ci}, some neutral
mesons~\cite{ref:Carosi-1990, ref:Schwingenheur-1995,
  ref:Abouzaid-2011, Kostelecky:1996fk, Colladay:1995qb,
  Colladay:1994cj, Kostelecky:1997mh, Isgur:2001yz,
  Kostelecky:2010bk}, muon~\cite{Bluhm:1999dx},
neutrino~\cite{Kostelecky:2003xn, Kostelecky:2003cr},
neutron~\cite{Cane:2003wp}, photon~\cite{Kostelecky:2008be}, Hydrogen
atom~\cite{Bluhm:1998rk} as well as some space based
experiments~\cite{Bluhm:2003un}. A summary of results of such studies
can be found in Ref~\cite{Kostelecky:2008ts}.  The best test of $CPT$
inavraiance has come from polarization studies of cosmic microwave
background radiation~\cite{Kostelecky:2008ts}.  In all these studies
there is no concrete indication of any breakdown of the $CPT$
invariance. However, if there is even an extremely small violation of
$CPT$, it would have very significant theoretical ramifications in
various models of new physics. If $CP$ violation is present in the
decay mode, it might overshadow the signature of $CPT$ violation in
the Dalitz plots. Therefore, usage of Dalitz plot for obervation of
$CPT$ violation must be dealt with deftly.  Nevertheless, probing
violations of Bose and $CPT$ symmetries by new methods is of paramount
importance. As was shown in Refs.~\cite{ref:Greenberg-2002,
  Colladay:1996iz} $CPT$ violation invariably leads to an associated
violation of Lorentz invariance in an interacting field theory. Though
very alluring, we do not dwell upon any signatures of Lorentz
violation in the Dalitz plot; as this is outside the scope of this
paper.

In this paper we shall point out methods, in search for the violations of Bose
symmetry and $CPT$ symmetry, in some three-body meson decays via the Dalitz
plot. This is in continuation of our efforts to use Dalitz plot as an
experimental tool to search for violations of some of the fundamental symmetries
in Nature, such as the $CP$ symmetry~\cite{Sahoo:2013mqa}. In this work we shall
analyze the observational signatures of violation of Bose symmetry and $CPT$
symmetry in the Dalitz plot. On the way we shall elucidate the techniques by
considering a few decay modes in which these searches would be fruitful.
Finally, we shall introduce the concept of and explain the utility of the Dalitz
`prism' which in its simplest form can be realized as a stacked up pile of
numerous Dalitz plots with increasing center-of-momentum energy. We conclude
emphasizing the importance of these new methods.

\begin{figure}[hbtp]
\centering
\includegraphics[scale=0.8]{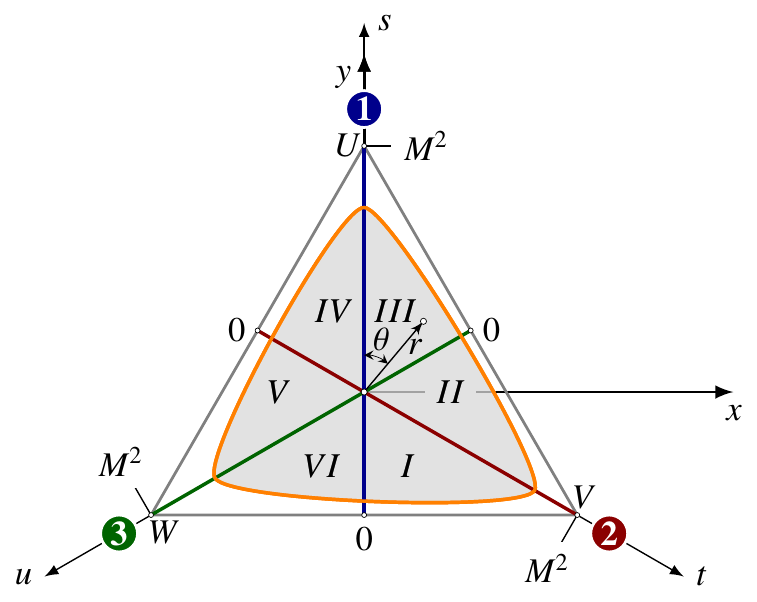}
\caption{(Color online) A hypothetical Dalitz plot with
  Mandelstam-like variables $s,t,u$ for the decay $X \to 1+2+3$. The
  three sides of the equilateral triangle $\triangle UVW$ are given by
  $s=0$, $u=0$ and $t=0$. At the three vertices we have $s = M^2$, $t
  = M^2$ and $u = M^2$. The three vertices in terms of the barycentric
  rectangular coordinate $(x,y)$ are given by $U=(0,2)$,
  $V=(\sqrt{3},-1)$ and $W = (-\sqrt{3},-1)$. For the barycentric
  polar coordinates $(r,\theta)$, the angle $\theta$ is measured from
  the vertical axis. The blobs with $\mathbf{1}$, $\mathbf{2}$ and
  $\mathbf{3}$ serve as mnemonic to suggest that the exchanges $s
  \leftrightarrow t \leftrightarrow u$ are equivalent to the particle
  exchanges $1 \leftrightarrow 2 \leftrightarrow 3$ respectively. The
  sextants of the Dalitz plot are also shown.}
\label{fig:Dalitz-plot-Mandelstam}
\end{figure}

Let us consider a general three-body decay process, say $ X \to 1 + 2
+ 3$, where the 4-momentum of the particle $i$ ($i \in \{ X,1,2,3 \}$)
is denoted by $p_i$ and its corresponding mass is denoted by $m_i$.
Let us also define the following Mandelstam-like variables: $s = (p_2
+ p_3)^2 = (p_X - p_1)^2$, $t = (p_1 + p_3)^2 = (p_X - p_2)^2$ and $u
= (p_1 + p_2)^2 = (p_X - p_3)^2$. It is well known that $s + t + u =
m_X^2 + m_1^2 + m_2^2 + m_3^2 = M^2$ (say). We can always construct a
ternary plot (see Fig.~\ref{fig:Dalitz-plot-Mandelstam}) of which
$(t,u,s)$ form the cartesian coordinates. The ternary plot can also be
described by a barycentric rectangular coordinate system $(x,y)$ or a
barycentric polar coordinate system $(r,\theta)$. For the polar
coordinate system the pole is at the centroid of the equilateral
triangle of the ternary plot and the polar axis passes through the
vertex for which $(t,u,s)=(0,0,M^2)$. It is quite straight-forward to
express the variables $s,t,u$ in terms of $r$, $\theta$ and $x,y$ as
follows:
\begin{align}
s &= \frac{M^2}{3} \Big( 1 + r \, \cos\theta \Big) = \frac{M^2}{3}
\Big( 1 + y \Big),\\
t &= \frac{M^2}{3} \left( 1 + r \, \cos\left( \frac{2\pi}{3} -
\theta \right) \right) = \frac{M^2}{6} \Big( 2 - \sqrt{3} x - y \Big),\\
u &= \frac{M^2}{3} \left( 1 + r \, \cos\left( \frac{2\pi}{3} +
\theta \right) \right) = \frac{M^2}{6} \Big( 2 + \sqrt{3} x - y \Big).
\end{align}

We know that $s,t,u$ take values in the following ranges: $(m_2 +
m_3)^2 \leqslant s \leqslant (m_X - m_1)^2$, $(m_1 + m_3)^2 \leqslant
t \leqslant (m_X - m_2)^2$, $(m_1 + m_2)^2 \leqslant u \leqslant (m_X
- m_3)^2.$ So when the final particles are ultra-relativistic, the
Dalitz plot tends to occupy the full region inside the equilateral
triangle of Fig.~\ref{fig:Dalitz-plot-Mandelstam}. In all cases the
Dalitz plot is inscribed inside the equilateral triangle. The density
of events inside the Dalitz plot is a consequence of the dynamics
driving the decay. Mathematically, if $A(r,\theta)$ is the amplitude
of the decay under consideration, the Dalitz plot density
$D(r,\theta)$ is directly proportional to $\modulus{A(r,\theta)}^2$.
If the full Dalitz plot can be constructed (i.e.\ if $0 \leqslant
\theta \leqslant 2\pi$), then the decay amplitude $A(r,\theta)$ can be
expanded in terms of a Fourier series as follows:
\begin{equation}
 A(r,\theta) = \sum_{n=0}^{\infty} \Big( S_{\!n} (r) \, \sin(n\theta)
+ C_n (r) \, \cos(n\theta) \Big),
\end{equation}
where $S_{\!n} (r)$ and $C_n (r)$ are the Fourier coefficients. It
would be profitable for us to divide the Dalitz plot into six sectors
or sextants by the medians of the equilateral triangle as shown in
Fig.~\ref{fig:Dalitz-plot-Mandelstam}.

It is now easy to explain the idea of observing the violations of Bose
symmetry and $CPT$ symmetry in the Dalitz plot. We shall first discuss
the Bose symmetry part. If particles $2$ and $3$ were identical
mesons, then the final state must remain symmetric under their
exchange as demanded by Bose symmetry. This implies that the Dalitz
distribution should remain symmetric under the exchange $t
\leftrightarrow u$. The decay amplitude $A(r,\theta)$ can also be
written as $A(t,u)$, such that $A(r,-\theta) \equiv A(u,t)$. If the
two particles $2$ and $3$ are not exactly identical, then the Bose
symmetry would not be strictly obeyed. In such a case, we can split
the decay amplitude into a part which is symmetric under $t
\leftrightarrow u$ exchange and another part which is non-symmetric
under the same exchange:
\begin{equation}
A(t,u) = A^S+A^N,
\end{equation}
where
\begin{align}
A^S &\equiv \frac{1}{2} \left( A(t,u) + A(u,t) \right) =
\sum_{n=0}^{\infty} C_n (r) \, \cos(n\theta),\\
A^N &\equiv \frac{1}{2} \left( A(t,u) - A(u,t)  \right) =
\sum_{n=0}^{\infty} S_{\!n} (r) \, \sin(n\theta).
\end{align}
The density of events in the Dalitz plot is proportional to the
amplitude mod-square. Only the interference term which is proportional
to $\Re\left( A^S \cdot A^{N*} \right)$ would give rise to an
asymmetry in the Dalitz plot under $t \leftrightarrow u$
exchange. Thus, we need the full Dalitz plot in this case. This can be
easily obtained if we construct the Dalitz plot from those events in
which particles $2$ and $3$ decay into different and distinct final
states. For example, the following decay modes
\begin{align*}
\left( \Kp, \Dp, \Dp_s \right) & \to
\underbrace{\pip(p_1)}_{\mu^+\nu_\mu} \;
\underbrace{\piz(p_2)}_{e^+ e^- \gamma} \;
\underbrace{\piz(p_3)}_{\gamma\gamma} \; , \\
\left( \Kp, \Dp,\Dp_s \right) & \to
\underbrace{\pim(p_1)}_{\mu^-\bar{\nu}_\mu} \;
\underbrace{\pip(p_2)}_{e^+\nu_e} \;
\underbrace{\pip(p_3)}_{\mu^+\nu_\mu} \; ,
\end{align*}
can be used for such a Bose symmetry violation study, since the
particles with 4-momenta $p_2$ and $p_3$ are the same but are
reconstructed from different final states. The extent of departure
from Bose symmetry can be quantified by using the conventional
left-right asymmetry of the Dalitz plot.

It is also possible to analyze three-body decays in which all the
final states are identical mesons, such that the final state is fully
Bose symmetric under the exchange of any two particles in it. Such a
situation would demand invariance under the exchange $s
\leftrightarrow t \leftrightarrow u$. This would imply that all the
sextants of the Dalitz plot would be symmetrical to one another when
we go from one to the other. Thus, if all the three final particles
are reconstructed from identical final states we would be left with
only one of the sextants of the Dalitz plot. For the Bose symmetry
test we need to have more than one sextant in our Dalitz plot. For
this we reconstruct two particles, say $1$ and $2$, from identical
final states and particle $3$ from different final state.  In this
case we would have sextants $VI$, $I$ and $II$ (or equivalently $III$,
$IV$ and $V$) in our Dalitz plot. If particles $2$ and $3$ are
identical bosons then these sextants should map from one to the
other. Any asymmetry among these sextants would be a signature of Bose
symmetry violation. Decay modes such as $(\eta, K^0_L, D^0) \to
\underbrace{\pi^0 (p_1)}_{\gamma\gamma} \underbrace{\pi^0
  (p_2)}_{\gamma\gamma} \underbrace{\pi^0 (p_3)}_{e^+ e^- \gamma}$,
$B^0 \to \underbrace{K^0_S (p_1)}_{\pi^+ \pi^-} \underbrace{K^0_S
  (p_2)}_{\pi^+ \pi^-} \underbrace{K^0_S (p_3)}_{\pi^0 \pi^0}$, can be
profitably used to search for the Bose symmetry violations in their
Dalitz plots.

The invariance under $CPT$ is a characteristic feature of any Lorentz
invariant local quantum field theory. Thus, it applies equally well to
both electro-weak and strong interactions. In weak interaction,
however, $CP$ violation is observed, which, as emphasized in the
introduction, can make the signature of $CPT$ violation unextractable
from the Dalitz plot. Keeping this in mind, we consider only those
decay modes which can occur via electromagnetic and strong
interactions, and thereby have no contribution from $CP$ violation in
them. $CP$ violation might still occur below the current experimental
bounds in these modes and mimic the signal for possible $CPT$
violation. It is, nevertheless, extremely interesting to look for any
unexpected violation of $CP$ or $CPT$ in strong or electromagnetic
interactions.  A nice example of such a process, free from $CP$
violation, is the decay modes $J/\psi \to N\pip\pim$, where $N$ can be
any of the following: $\piz$, $\omega$, $\eta$, $\phi$.  The amplitude
$A(r,\theta)$ for the process $J/\psi \to N \pip \pim$ can be expanded
in a Fourier series as follows:
\begin{equation}
\label{eq:Art}
A(r,\theta) = \sum_{n=0}^{\infty} \Big( s_{n} (r) \, \sin(n\theta)
+ c_{n} (r) \, \cos(n\theta) \Big),
\end{equation}
where $s_{n} (r)$ and $c_{n} (r)$ are Fourier coefficients which are
in general complex.  Under $CPT$ the angle $\theta$ goes to $-\theta$
and the complex Fourier coefficients ($s_n (r)$ and $c_n (r)$)
transform to their respective complex conjugates ($s_n^* (r)$ and
$c_n^* (r)$).  Therefore, $CPT$ invariance implies that $A(r,\theta) =
A^*(r,-\theta)$. Moreover, for a self-conjugate process the initial
and final state must have the same $CP$ if $CP$ is conserved. Hence,
conservation of $CP$ and $CPT$ jointly implies that all the $s_n (r)$
are zero and all the $c_n (r)$ are purely real. This restricts the
amplitude in Eq.~\eqref{eq:Art} and the Dalitz plot density to be
symmetric under $\theta\leftrightarrow -\theta$. If $CP$ is conserved,
any asymmetry in the Dalitz plot under $\theta\leftrightarrow -\theta$
would therefore be a signature of $CPT$ violation as discussed below.

The amplitude $\bar A(r,-\theta)$ for the $CP$ conjugate process,
assuming $CPT$ violation, is given by:
\begin{equation}
\label{eq:Artb}
\bar A(r,-\theta) = \sum_{n=0}^{\infty} \Big( - \bar s_{n} (r) \,
\sin(n\theta) + \bar c_{n} (r) \, \cos(n\theta) \Big),
\end{equation}
where $\bar s_n (r)$ and $\bar c_n (r)$ are Fourier coefficients that
are complex and are necessarily different from $s_n^* (r)$ and $c_n^*
(r)$ respectively unless $CPT$ is conserved~\cite{footnote}.  The
coefficients $s_n (r)$, $c_n (r)$, $\bar s_n (r)$ and $\bar c_n (r)$
can be written as
\begin{align*}
s_n (r) &= \left( \modulus{s_n (r)} + \epsilon_n^s (r) \right)
e^{i\delta_n^s}, &\bar s_n (r) = \left( \modulus{s_n (r)} -
\epsilon_n^s (r) \right) e^{i\delta_n^s},\\
c_n (r) &= \left( \modulus{c_n (r)} + \epsilon_n^c (r) \right)
e^{i\delta_n^c}, &\bar c_n (r) = \left( \modulus{c_n (r)} -
\epsilon_n^c (r) \right) e^{i\delta_n^c},
\end{align*}
where, $\delta_n^{s,c}$ are the strong phases and $\epsilon_n^{s,c}
(r)$ are $CPT$ violating terms, i.e.\ for the case of $CPT$ invariance
they vanish identically. Since we have assumed that $CP$ is conserved
no explicit weak phase dependence is retained in $s_n (r)$, $c_n (r)$,
$\bar s_n (r)$ and $\bar c_n (r)$.

Since in our case the process and its $CP$ conjugate process are the
same, the amplitude which comes into picture is the average of both
$A(r,\theta)$ and $\bar A(r,-\theta)$:
\begin{align}
A &\equiv \frac{1}{2}\Big( A(r,\theta) + \bar A(r,-\theta)\Big) \nn\\
\label{eq:Amp}
&=\sum_{n=0}^{\infty} \Big( \epsilon_n^s (r) \, \sin(n\theta)
e^{i\delta_n^s}+ \modulus{c_n (r)} \, \cos(n\theta) e^{i\delta_n^c}
\Big).
\end{align}
The logic for the average is easy to realize by observing that if
$CPT$ is conserved, $\epsilon_n^{s,c} (r)=0$ and the amplitude in
Eq.~\eqref{eq:Amp} reduces to that in Eq.~\eqref{eq:Art} as
expected. The Dalitz distribution is proportional to $\modulus{A}^2$
and any asymmetry under $\theta \leftrightarrow -\theta\equiv t
\leftrightarrow u$ can arise only from the term odd under $\theta$
which is proportional to
\begin{equation*}
\sum_{n,m=0}^{\infty} \modulus{c_n (r)} \epsilon_m^s (r) \cos\left(
\delta_n^c - \delta_m^s \right) \cos(n\theta) \sin(m\theta).
\end{equation*}
This interference term survives only if $CPT$ is violated. We have
thus demonstrated mathematically how $CPT$ violation leads to
asymmetry in the Dalitz plot. It should be noted that the observation
of such an asymmetry would be an unambiguous signature of $CPT$
violation (or $CP$ violation in strong and electromagnetic
interaction) and would demand the presence of $CPT$ (or $CP$)
violating new physics. The usual left-right Dalitz plot asymmetry can
be used to quantify this asymmetry in the Dalitz plot. It is possible
to look for $CPT$ violation in any self-conjugate process of the form
$X \to N M \bar M$ which proceeds via strong or electromagnetic
interactions preserving $CP$ in the decay.

\begin{figure}
\centering
\includegraphics[scale=0.8]{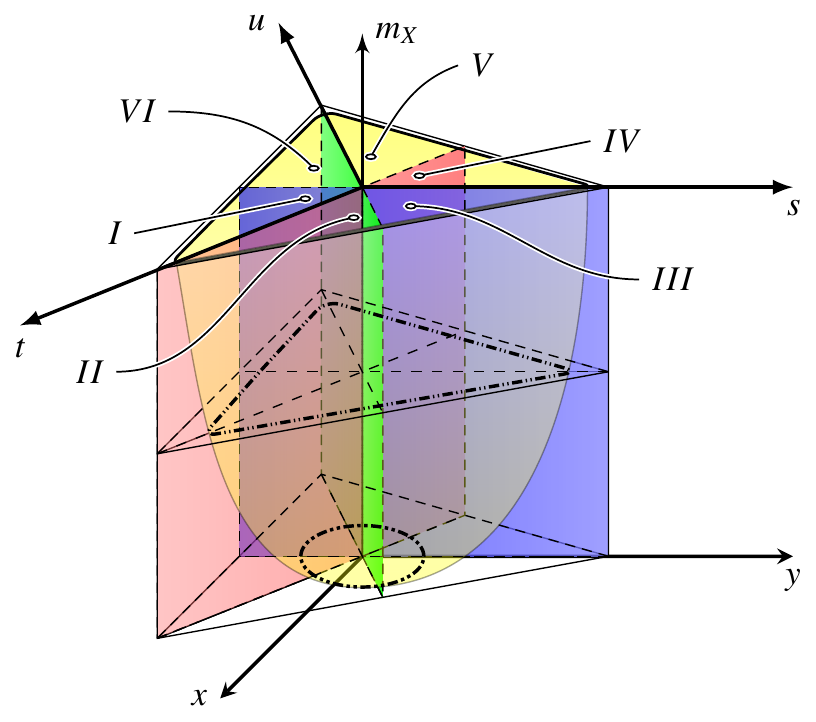}
\caption{(Color online) The schematic figure of the Dalitz `prism'.
  To get the Dalitz plot corresponding to a certain value of $m_X$ one
  simply slices the Dalitz `prism' parallel to the $xy$ plane at that
  particular value of $m_X$. The sextants of the Dalitz plots are
  subsumed into the six analogously labeled identical wedges of the
  Dalitz prism.}
\label{fig:Dalitz-prism}
\end{figure}

The Bose and $CPT$ symmetries are expected to hold firmly. Their violations, if
any, would by virtue be extremely small. In order to possibly observe such tiny
numbers, one would require as large a sample of events as possible. In purview
of this {\em a new concept} of Dalitz `prism' is developed here. So far in the
discussions on Bose and $CPT$ symmetries, details of the initial particle $X$
played no role. In fact the particle $X$ can be replaced by, say $e^+ e^-$, such
that $m_X$ denotes the total energy in the center-of-momentum frame. In such a
situation we are dealing with continuum production of particles $1$, $2$ and
$3$, e.g.\ $e^+ e^- \to \pi^+ \pi^- \pi^0$. Considering such continuum
productions in association with the decays of several resonances would provide
significantly larger statistics to study the violations of Bose and $CPT$
symmetry. To facilitate such a study we note that the Dalitz plot can be
generalized into a three-dimensional plot, which we call as the Dalitz `prism'
(Fig.~\ref{fig:Dalitz-prism}). This prism is a regular right triangular prism.
For a given value of $m_X$ one can slice this `prism' to obtain the Dalitz plot.
Decay events corresponding to all possible values of $m_X$ fill up only those
regions of the `prism' which are allowed by conservation of energy and momentum.
This idea of Dalitz prism can be extended to include cases such as $X(p_X)
\to N(p_1) M(p_2) \bar M(p_3)$ where $N$ can represent more than one particle
and $p_1$ is, therefore, the total 4-momentum of all those particles denoted by
$N$. One example of such a mode is $X\to N\pip\pim$ where the initial state $X$
can be a resonance such as $J/\psi$ or even $e^+ e^-$, the final state $N$ can
be $\Kp\Km$, $\piz\Kp\Km$, $\Kp\Km\eta$, $\omega\piz$, $p\bar p$, $p \bar p
\piz$ and $n\bar n$.  In such a case, the value of $p_1^2$ is not fixed even
though for a given initial state configuration $p_X^2 = m_X^2$ is fixed at a
constant value. One can also vary both $p_1^2=m_1^2$ as well as $m_X^2$, such as
when $X=e^+ e^-$. For such cases we can again construct a prism whose $z$ axis
denotes $M^2 = m_X^2 + m_1^2 + m_2^2 + m_3^2$, such that the $M^2$ value can
vary even if either $m_X^2$ or $m_1^2$ or both vary. The $xy$-plane of the prism
is spanned by the various values of $s$, $t$ and $u$ as before. When $p_X$,
$p_2$ and $p_3$ are precisely measured, $p_1$ need not be measured, as $p_1 =
p_X - p_2 - p_3$ from conservation of 4-momentum. Similarly, one need not
measure $p_X$ when $p_1$, $p_2$, and $p_3$ are precisely measured.  All the
events allowed by conservation of energy and 3-momentum populate the interior of
this general prism. Even though, slices of this prism do not give any Dalitz
plot, because the recorded events are no longer just three-body decays, we shall
nevertheless refer to it as Dalitz prism as well. The Dalitz prism can,
therefore, subsume all cases where $m_X^2$ and/or $m_1^2$ varies. The
distribution of events on the $z$ axis is irrelevant for our discussion. One
only needs to take a projection of all the events recorded in this unified prism
onto its base and look for asymmetry in the resulting triangular plot. Usage of
the Dalitz `prism' as explained above, thus liberates the methods discussed here
from the shackles of branching fractions and thereby enhances the sensitivity of
the search for violations of Bose and $CPT$ symmetries. It is noteworthy that
the use of Dalitz `prism' in the study of $CP$ violation~\cite{Sahoo:2013mqa}
can also be advantageous. The Dalitz `prism' in its generalized form is hence a
very significant tool to study the fundamental symmetries of nature using
multibody decays.

We have thus shown how Bose symmetry and $CPT$ symmetry violations can lead to
asymmetries in the Dalitz plot. We have also developed the new concept of Dalitz
`prism' which can be used to gather the huge statistics needed for an effective
search for the Bose and $CPT$ symmetry violations from studies of meson decays.
Since both Bose symmetry and $CPT$ symmetry are of fundamental importance to
foundations of modern field theory, it is worthwhile to check their validity in
the realm of unstable and composite particles such as the mesons.

\acknowledgments

This work is supported in part by DOE under Contracts No.
DE-FG02-96ER40969ER41155 and No. DE-FG02-04ER41291. NGD thanks Arjun
Menon for reading the manuscript. We are thankful to Alan
Kostelecky for going through the initial version of the manuscript and
giving  valuable suggestions.


\begin{thebibliography}{99}

\bibitem{ref:Bose-1924} S.N. Bose, \textit{Zeitschrift f\"{u}r Physik}
\textbf{26}, 178 (1924).

\bibitem{ref:Fermi-1926} E. Fermi, \textit{Rendiconti Lincei}
\textbf{3}, 145 (1926).

\bibitem{ref:Hilborn} R.C. Hilborn, \textit{Bull.\ Am.\ Phys.\ Soc.}
  \textbf{35}, 982 (1990).

\bibitem{ref:Tino-1994} G.M. Tino, \textit{Il Nuovo Cimento D}
  \textbf{16}, 523 (1994).

\bibitem{ref:Tino-2000} G.M. Tino, \textit{AIP Conf.\ Proc.}
  \textbf{545}, 260 (2000).
  
\bibitem{ref:Modugno-1998} G. Modugno, M. Inguscio and G.M. Tino,
\textit{Phys.\ Rev.\ Lett.} \textbf{81}, 4790 (1998).
  
\bibitem{ref:Hilborn-1996} R.C. Hilborn and C.L. Yuca,
\textit{Phys.\ Rev.\ Lett.} \textbf{76}, 2844 (1996).

\bibitem{ref:Angelis} M. de Angelis, G. Gagliardi, L. Gianfrani, and
G.M. Tino, \textit{Phys.\ Rev.\ Lett.} \textbf{76}, 2840 (1996).

\bibitem{ref:Fivel-1991} D.I. Fivel, \textit{Phys.\ Rev.\ A}
  \textbf{43}, 4913 (1991). Objections to the bound on violations of
  Bose statistics for photons given in this paper have been raised;
  see O. W. Greenberg in \textit{Workshop on Harmonic Oscillators},
  NASA Conference Pub. 3197, edited by D. Han, Y. S. Kim and
  W. W. Zachary (NASA, Greenbelt, 1993).


\bibitem{ref:Gerry-1997} C.C. Gerry and R.C. Hilborn,
  \textit{Phys.\ Rev.\ A} \textbf{55}, 4126 (1997).

\bibitem{ref:manko} V.I. Man'ko and G.M. Tino,
  \textit{Phys.\ Lett.\ A} \textbf{202}, 24 (1995).

\bibitem{ref:Ignatiev} A. Yu. Ignatiev, G.C. Joshi and M. Matsuda,
  \textit{Mod.\ Phys.\ Lett.\ A} \textbf{11}, 871 (1996).

\bibitem{ref:DeMille-1999} D. DeMille, D. Budker, N. Derr and E.
  Deveney, \textit{Phys.\ Rev.\ Lett}. \textbf{83}, 3978 (1999).

\bibitem{ref:DeMille-2000} D. DeMille, D. Budker, N. Derr and E.
  Deveney \textit{AIP Conf.\ Proc.} \textbf{545}, 227 (2000).

\bibitem{ref:Greenberg-1988um} O.W. Greenberg and R.N. Mohapatra,
  \textit{Phys.\ Rev.\ D} {\bf 39}, 2032 (1989).
  
\bibitem{ref:Mills} A.P. Mills and P. Zukerman,
  \textit{Phys.\ Rev.\ Lett.} \textbf{64}, 2637 (1990).

\bibitem{ref:Gidley} D. Gidley \textit{et al.},
  \textit{Phys.\ Rev.\ Lett.} \textbf{66}, 1302 (1991).

\bibitem{ref:Asai} A. Asai \textit{et al.}, \textit{Phys.\ Rev.\
Lett.} \textbf{66}, 1298 (1991).

\bibitem{ref:OPAL} OPAL Collab. (M.Z. Akrawy \textit{et al.}),
  \textit{Phys.\ Lett.} \textbf{257}, 531 (1991).
  
\bibitem{ref:English-2010} D. English, V. Yashchuk and D. Budker,
  \textsf{arXiv:1001.1771}.

\bibitem{ref:Kozlov-2012ag} G.A. Kozlov, \textsf{arXiv:1210.7472
  [hep-ph].}
  
\bibitem{ref:Gninenko-2011ws} S.N. Gninenko, A.Y. Ignatiev and
  V.A. Matveev, \textit{Int.\ J.\ Mod.\ Phys.\ A} {\bf 26}, 4367
  (2011).

\bibitem{ref:Jackson-2008bs} M.G. Jackson, \textit{Phys.\ Rev.\ D}
  {\bf 77}, 127901 (2008).

\bibitem{ref:Jackson-2008xq} M.G. Jackson, \textit{Phys.\ Rev.\ D}
  {\bf 78}, 126009 (2008).
  
\bibitem{ref:Greenberg-1991av} O.W. Greenberg, MDDP-PP-91-285.
  
\bibitem{ref:Greenberg-1991au} O.W. Greenberg, in ``Moscow 1991,
  Proceedings, Sakharov memorial lectures in physics, vol. 1'' 29-37
  and Maryland Univ. College Park - PP 92-006 (91,rec.Aug.) 9 p

\bibitem{ref:Greenberg-2000zy} O.W. Greenberg, \textit{AIP
  Conf.\ Proc.} {\bf 545}, 113 (2004).

\bibitem{ref:Belinfante} F. J. Belinfante, \textit{Physica}
  \textbf{6}, 870 (1939).

\bibitem{ref:Pauli-Belinfante} W. Pauli, F. J. Belinfante,
  \textit{Physica} \textbf{7}, 177 (1940).

\bibitem{ref:Pauli-1940} W. Pauli, \textit{Phys. Rev.} \textbf{58},
  716 (1940).

\bibitem{ref:Schwinger-1951} J. Schwinger, \textit{Phys. Rev.}
  \textbf{82}, 914 (1951).

\bibitem{ref:Schwinger-1953} J. Schwinger, \textit{Phys. Rev.}
  \textbf{91}, 713 (1953).

\bibitem{ref:Luders-1954} G. L\"{u}ders, \textit{Kgl. Danske
  Videnskab. Selskab, Mat.-fys. Medd.} \textbf{28}, No. 5 (1954)

\bibitem{ref:Bell-1955} J. S. Bell, \textit{Proc. Roy. Soc. (London)}
  \textbf{A231}, 79 (1955).

\bibitem{ref:Pauli-1955} W. Pauli, ``Exclusion Principle, Lorentz
  Group and Reflection of Space-Time and Charge'', in \textit{Niels
    Bohr and the Development of Physics}, ed. W. Pauli, L. Rosenfeld
  and V. Weisskopf (McGraw-Hall Book Company, Inc., New York, and
  Pergamon Press, Inc., London, 1955)

\bibitem{ref:Luders-1957} G. L\"{u}ders, \textit{Ann. Phys.}
  \textbf{2}, 1 (1957).

\bibitem{ref:Jost-1957}R.~Jost, \textit{Helv.\ Phys.\ Acta}
  \textbf{30}, 409 (1957).
  
\bibitem{ref:Jost-1960}R.~Jost, in \textit{Theoretical Physics in
Twentieth Century} (Interscience, New York, 1960).

\bibitem{ref:Streater-Wightman} R. F. Streater and A. S. Wightman,
\textit{PCT, Spin \& Statistics, and All That} (Benjamin, New York,
1964).

\bibitem{ref:Fainberg-1969} V.~Ya.~Fainberg, \textit{Sov.\ Phys.\
Usp.\ } \textbf{11}, 506 (1969).

\bibitem{ref:Fonda-Ghirardi} L.~Fonda and G.~C.~Ghirardi,
\textit{Symmetry Principles in Quantum Theory} (Dekker, New York,
1970).

\bibitem{ref:Haag-1996} R.~Haag, \textit{Local Quantum Physics}
(Springer, Berlin, 1996).


\bibitem{ref:Luders-1957pr} G. L\"{u}ders, B. Zumino, \textit{Phys.
  Rev.} \textbf{106}, 385 (1957).

\bibitem{ref:Luders-1958} G. L\"{u}ders, B. Zumino, \textit{Phys.
  Rev.} \textbf{110}, 1450 (1958).

\bibitem{ref:streater-1980} R.~F.~Streater and A.~S.~Wightman (1980),
  \textit{$PCT$, spin and statistics, and all that}. Princeton, NJ:
  Princeton Univ. Press.

\bibitem{ref:Greenberg-2002} O.~W.~Greenberg,
  \textit{Phys.\ Rev.\ Lett.\ } \textbf{89}, 231602 (2002).


\bibitem{Dalitz:1990md} R.~H.~Dalitz, Nucl.\ Phys.\ Proc.\ Suppl.\ 
{\bf 24A}, 3 (1991).


\bibitem{ref:Oksak-1968} A.~I.~Oksak, and I.~T.~Todorov,
\textit{Comm.\ Math.\ Phys.\ } \textbf{11}, 125 (1968).

\bibitem{ref:Bogoliubov-1975} N. N. Bogoliubov, A. A. Logunov, and I.
T. Todorov, \textit{Introduction to Axiomatic Quantum Field Theory}
(Benjamin, Reading, 1975).

\bibitem{Majorana:1932rj} E.~Majorana, Nuovo Cim.\  {\bf 9}, 335
(1932).

\bibitem{ref:Nambu-1966} Y.~Nambu, \textit{Prog.\ Theor.\ Phys.\
Suppl.\ } \textbf{37} \& \textbf{38}, 368 (1966).

\bibitem{ref:Nambu-1967} Y.~Nambu, \textit{Phys.\ Rev.\ }
\textbf{160}, 1171 (1967).

\bibitem{ref:Okubo-1958} S. Okubo, \textit{Phys. Rev.} \textbf{109},
  984 (1958).

\bibitem{ref:Okubo-1962} S. Okubo, \textit{Phys. Rev.} \textbf{128},
  1921 (1962).

\bibitem{Bluhm:1997ci} R.~Bluhm, V.~A.~Kostelecky and N.~Russell,
  \textit{Phys.\ Rev.\ Lett.} {\bf 79}, 1432 (1997).

\bibitem{ref:Carosi-1990} R. Carosi \textsl{et al.},
  \textit{Phys. Lett. B} \textbf{237}, 303 (1990).

\bibitem{ref:Schwingenheur-1995} B. Schwingenheur \textsl{et al.},
  \textit{Phys. Rev. Lett.} \textbf{74}, 4376 (1995).

\bibitem{ref:Abouzaid-2011} E. Abouzaid \textsl{et al.},
  \textit{Phys. Rev. D} \textbf{83}, 092001 (2011).

\bibitem{Kostelecky:1996fk} V.~A.~Kostelecky and R.~J.~Van Kooten,
  \textit{Phys.\ Rev.\ D} {\bf 54}, 5585 (1996).

\bibitem{Colladay:1995qb} D.~Colladay and V.~A.~Kostelecky,
  \textit{Phys.\ Rev.\ D} {\bf 52}, 6224 (1995).

\bibitem{Colladay:1994cj} D.~Colladay and V.~A.~Kostelecky,
  \textit{Phys.\ Lett.\ B} {\bf 344}, 259 (1995).

\bibitem{Kostelecky:1997mh} V.~A.~Kostelecky,
  \textit{Phys.\ Rev.\ Lett.}  {\bf 80}, 1818 (1998).

\bibitem{Isgur:2001yz} N.~Isgur, V.~A.~Kostelecky and
  A.~P.~Szczepaniak, \textit{Phys.\ Lett.\ B} {\bf 515}, 333 (2001).

\bibitem{Kostelecky:2010bk} A.~Kostelecky and R.~Van Kooten,
  \textit{Phys.\ Rev.\ D} {\bf 82}, 101702 (2010).

\bibitem{Bluhm:1999dx} R.~Bluhm, V.~A.~Kostelecky and C.~D.~Lane,
  \textit{Phys.\ Rev.\ Lett.}  {\bf 84}, 1098 (2000).

\bibitem{Kostelecky:2003xn} V.~A.~Kostelecky and M.~Mewes,
  \textit{Phys.\ Rev.\ D} {\bf 70}, 031902 (2004).

\bibitem{Kostelecky:2003cr} V.~A.~Kostelecky and M.~Mewes,
  \textit{Phys.\ Rev.\ D} {\bf 69}, 016005 (2004).

\bibitem{Cane:2003wp} F.~Cane, D.~Bear, D.~F.~Phillips, M.~S.~Rosen,
  C.~L.~Smallwood, R.~E.~Stoner, R.~L.~Walsworth and V.~A.~Kostelecky,
  \textit{Phys.\ Rev.\ Lett.}  {\bf 93}, 230801 (2004).

\bibitem{Kostelecky:2008be} V.~A.~Kostelecky and M.~Mewes,
  \textit{Astrophys.\ J.}  {\bf 689}, L1 (2008).

\bibitem{Bluhm:1998rk} R.~Bluhm, V.~A.~Kostelecky and N.~Russell,
  \textit{Phys.\ Rev.\ Lett.} {\bf 82}, 2254 (1999).

\bibitem{Bluhm:2003un} R.~Bluhm, V.~A.~Kostelecky, C.~D.~Lane and
  N.~Russell, \textit{Phys.\ Rev.\ D} {\bf 68}, 125008 (2003).

\bibitem{Kostelecky:2008ts} V.~A.~Kostelecky and N.~Russell,
  \textit{Rev.\ Mod.\ Phys.}  {\bf 83}, 11 (2011).

\bibitem{Colladay:1996iz} D.~Colladay and V.~A.~Kostelecky,
\textit{Phys.\ Rev.\ D} {\bf 55}, 6760 (1997).

\bibitem{Sahoo:2013mqa} D.~Sahoo, R.~Sinha, N.~G.~Deshpande and
  S.~Pakvasa, \textit{Phys.\ Rev.\ D} \textbf{89}, 071903 (2014).


\bibitem{footnote} We have retained only the leading contribution
  which is sufficient for establishing $CPT$ violation. For $CPT$
  violation, the Hamiltonian $\mathscr{H}$ driving the process under
  consideration does not commute with $CPT$, i.e.\ $\left( CPT \right)
  \mathscr{H} \left( CPT \right)^{-1} = \bar{\mathscr{H}} \neq
  \mathscr{H}$, where $\bar{\mathscr{H}}$ is the Hamiltonian for the
  $CP$ conjugate process. If $CPT$ were conserved, only then
  $\bar{\mathscr{H}}=\mathscr{H}$.
\end{thebibliography}
\end{document}